\documentclass[prd,12pt,nofootinbib,preprint]{revtex4-1}

\pdfoutput=1
\usepackage[T1]{fontenc}
\usepackage[utf8]{inputenc}
\DeclareUnicodeCharacter{200E}{}
\usepackage{mathtools}
\usepackage{amsmath,amssymb}
\usepackage{amsfonts}
\usepackage{epsfig}
\usepackage{url}
\usepackage{subfigure}
\usepackage{graphicx}
\usepackage{grffile}
\usepackage[usenames,dvipsnames]{color}
\usepackage{slashed}
\usepackage{array}
\usepackage{multirow}
\usepackage[colorlinks,citecolor=blue]{hyperref}
\usepackage{color}
\usepackage{cancel}
\usepackage{gensymb}

\def\a{\alpha}

\def\b{\beta}
\def\g{\gamma}

\def\l{\lambda}

\def\s{\sigma}

\def\q2 {q^2}

\def\a {\alpha}
\def\b {\beta}
\def\g {\gamma}

\def\l {\lambda}

\def\bar {\overline}

\def\be {\begin{equation}}
\def\ee {\end{equation}}
\def\beq {\begin{equation}}
\def\eeq {\end{equation}}

\def\bra {\langle}
\def\ket {\rangle}

\newcommand{\besub}{\begin{subequations}}
\newcommand{\eesub}{\end{subequations}}
\newcommand{\bea}{\begin{eqnarray}}
\newcommand{\eea}{\end{eqnarray}}

\def\g{\Gamma}

\def\beq{\begin{equation}}
\def\eeq{\end{equation}}
\def\barr{\begin{array}}
\def\earr{\end{array}}

\begin{document}
\title{Impact of dark scalars on the radiative $H^+ W^- Z$ vertex}

\author{Nabarun Chakrabarty}
\email{chakran@iisc.ac.in, chakrabartynabarun@gmail.com}
\affiliation{Centre for High Energy Physics, Indian Institute of Science, C.V.Raman Avenue,
Bangalore 560012, India}

\begin{abstract}
An $H^+ W^- Z$ vertex, i.e, the interaction connecting a singly charged physical scalar to the 
$W^\pm,~Z$ gauge bosons is absent at the tree level in a scalar sector comprising only $SU(2)_L$ doublets. However, the interaction can be generated radiatively in such  a case, an example of which is a two-Higgs doublet model. In this study, we extend a two-Higgs doublet scenario by additional 
$SU(2)_L$ scalar doublets. Upon endowing with appropriate discrete symmetries, these additional doublets so introduced can furnish successful candidates of dark matter. Interestingly, the same "dark" scalars can also radiatively enhance the strength of the 
$H^+ W^- Z$ interaction. We compute the $H^+ W^- Z$ vertex at one-loop in a non-linear gauge that eliminates certain unphysical interactions. While all possible constraints are taken into account in doing so, particular emphasis is given to the ones stemming from dark matter. Correlations between the dark matter observables and the $H^+ W^- Z$ interaction-strength  are studied. The study thus connects a radiatively generated $H^+ W^- Z$ vertex to dark matter phenomenology.    
\end{abstract}	
\maketitle

 \section{Introduction}\label{intro}

The Higgs-discovery at the Large Hadron Collider (LHC) \cite{cms:2012ufa,Aad:2012tfa} completes the particle content of the Standard Model (SM). Moreover, the interaction strengths of the discovered boson with the SM fermions and gauge bosons are found to be in agreement with the corresponding SM values. Despite such a success, the SM model cannot be a complete theory since it is plagued with shortcomings both on theoretical as well as experimental fronts. On the experimental side, observation of galactic rotation curves, gravitational lensing and anisotropies in cosmic microwave background (CMB) advocate the existence of cosmologically stable dark matter (DM) in the present universe~\cite{Spergel:2006hy,Aghanim:2018eyx}. Assuming DM has a particle character, no such particle candidate(s) can be accommodated within the SM alone. This necessitates dynamics beyond the SM (BSM). A thermal DM particle that thermalises with the visible sector with a typical weak interaction strength is named a Weakly Interacting Massive Particle (WIMP) (see \cite{Roszkowski:2017nbc} for a review). It \emph{freezes-out} from the thermal bath as its interaction rate is surpassed by the Hubble expansion rate.

The lack of precise information on dark matter quantum numbers allows for the possibility that
DM consists of more than one type of particle. Such multiparticle DM frameworks allow for DM-DM interaction~(some recent studies are \cite{Biswas:2013nn,Bhattacharya:2013hva,Bian:2013wna,Esch:2014jpa,Bhattacharya:2016ysw,Ahmed:2017dbb,Poulin:2018kap,
Herrero-Garcia:2017vrl,Herrero-Garcia:2018qnz,
Aoki:2018gjf,Aoki:2017eqn,Elahi:2019jeo,
Bhattacharya:2019fgs,Biswas:2019ygr,DuttaBanik:2020jrj}). While such processes can contribute to the thermal relic, they do not have a role in DM-nucluon scattering rates. Thus, a multipartite DM model can evade the ever tightening bound on the
direct detection (DD) cross section while successfully accounting for the thermal relic. A multipartite DM framework comprising scalar DM candidates is obtained by augmenting the SM 
by additional scalars and introducing appropriate discrete symmetries in order to ensure DM stability. A popular scalar dark matter scenario is the
inert-doublet model (IDM) in which case the SM is augmented by an additional $SU(2)_L$ scalar doublet featuring a $\mathbb{Z}_2$ symmetry. In fact, an interesting multipartite extension of the IDM 
was proposed recently where the SM was augmented by two additional $SU(2)_L$ scalar doublets that were charged under $\mathbb{Z}_2 \times \mathbb{Z}^\prime_2$~\cite{Borah:2019aeq}.

A two-Higgs doublet model (2HDM) \cite{Branco:2011iw,Deshpande:1977rw} is a widely studied extension of the SM Higgs sector. Motivated by the minimal supersymmetric standard model (MSSM) in part, it potentially shuts off the flavour changing neutral currents (FCNC), furnishes additional sources of CP-violation that can eventually explain the observed matter-antimatter imbalance, and, poses a solution to the strong CP problem. In fact, a 2HDM is in fact the smallest
$SU(2)_L$ multiplet predicting a singly charged Higgs 
$H^+$. The more well known collider search channels of the same are its fermionic decays. The charged Higgs
is searched at the LHC through different production and decay modes. 
The $H^+ \to t \bar{b}$ decay is looked for in the search of a heavy $H^+$ whereas the preferred search channel for a light charged Higgs is the $H^+ \to \bar{\tau} 
\nu_\tau$ channel. However, such signals are generally swamped by a heavy QCD background. An alternate strategy therefore is to search for its bosonic decays 
$H^+ \to W^+ h, W^+ Z, W^+ \gamma$. The last two of the aforementioned modes are \emph{prima facie} more interesting since they arise only at one-loop in multi-Higgs doublet models\footnote{The $H^+ W^- Z$ interaction is present at the tree level itself in models with scalar triplets. Examples are the Georgi-Machacek model~\cite{Georgi:1985nv,Chanowitz:1985ug,Gunion:1989ci,Aoki:2007ah,Chiang:2012cn,Hartling:2014zca,Hartling:2014aga,Chiang:2014bia,Logan:2015xpa,Blasi:2017xmc,Cen:2018okf,Ghosh:2019qie,Banerjee:2019gmr} and scalar-triplet extensions of the MSSM~\cite{Bandyopadhyay:2014vma, Bandyopadhyay:2015ifm, Bandyopadhyay:2017klv}}. The absence of the 
$H^+ W^- Z$ coupling at the tree level is an artefact of the isospin symmetry of the kinetic terms of the Higgs sector. This symmetry is broken at one loop level
through effects from other sectors that do not respect the custodial invariance, these vertices are induced
at loop level. Momentum dependent interactions appear therein consequently.
The $H^+ W^- Z$ vertex indeed has stirred some interest in the past. Its strength has been estimated for the minimal supersymmetric standard model (MSSM)~\cite{Raychaudhuri:1992tm,Arhrib:2007rm}, a 
$\mathbb{Z}_2$ symmetric 2HDM \cite{Kanemura:1997ej,Kanemura:1999tg}, an aligned 2HDM \cite{Abbas:2018pfp} and a particular version of 3HDM containing two \textit{active} and one \textit{inert} doublet \cite{Moretti:2015tva}, and, the 2HDM in presence of a color-octet scalar isodoublet~\cite{Chakrabarty:2020msl}.

In this work, we consider a scenario where a 2HDM is augmented by two inert scalar doublets that are charged non-trivially under a $\mathbb{Z}_2 \times \mathbb{Z}^\prime_2$ symmetry. The ensuing DM phenomenology  is looked at and the parameter regions
predicting the requisite thermal relic  are identified. In particular, the role of DM-DM conversions in generating the observed relic in the IDM \emph{desert region} is illuminated. On the other hand, we note that the $SU(2)_L$
nature of the inert scalars makes them contribute radiatively to the $H^+ W^- Z$ vertex. That is, the contribution of the inert scalars adds to the one coming from the non-inert scalars 
coming from the pure 2HDM. We subsequently estimate the strength of the $H^+ W^- Z$ vertex for the parameter regions allowed by the DM constraints. In the process, we adopt the non-linear gauge to do away with unphysical vertices involving goldstones. In all, as a major upshot, this study correlates the $H^+ W^- Z$ interaction strength with the observables in the DM sector, the primary one being the thermal relic for this case.

The paper is organised as follows. We introduce the 2HDm + two inert doublet scenario in section \ref{model}. Section \ref{form_factor} elaborates on the calculation of the one-loop 
$H^+ W^- Z$ vertex in the chosen non-linear gauge. We discuss the relevant constraints in section \ref{constraints}. Section \ref{DM} elucidates the key features of the DM phenomenology and section
\ref{combined} aims to present results combining the DM phenomenology and the one-loop results for the 
$H^+ W^- Z$ vertex.
Finally we conclude in section \ref{conclusion}. Various important formulae are relegated to the Appendix. 

\section{A framework with two inert doublets}\label{model}
This study extends a 2HDM (comprising two scalar $SU(2)_L$ doublets $\phi_1$ and 
$\phi_2$) with two additional scalar $SU(2)_L$ doublets $\eta_1$ and 
$\eta_2$. An additional $\mathbb{Z}_2 \times \mathbb{Z}^\prime_2$ is introduced under which
the SM fermions and $\phi_{1,2}$ transform trivially. The charges of $\eta_{1,2}$ under the same are given in Table~\ref{qno}.

\begin{table}[htpb!]
\begin{center}\scalebox{1.82}{
\begin{tabular}{|c|c|c|}
\hline
Field & $\mathbb{Z}_2$ & $\mathbb{Z}^\prime_2$\\
\hline
$\eta_1$ & + & - \\
$\eta_2$ & - & + \\
\hline
\end{tabular}}
\end{center}
\caption{Discrete charges of $\eta_1$ and $\eta_2$.}
\label{qno}
\end{table} 

The most general scalar potential consistent with the gauge and discrete symmetries reads:
\besub
\bea
V_2 &=& - m_{11}^2 |\phi_1|^2 - m_{22}^2 |\phi_2|^2
+ m_{12}^2 (\phi_1^\dagger \phi_2 + \text{h.c.})
+ \mu_1^2 |\eta_1|^2 + \mu_2^2 |\eta_2|^2 \\
V_4^a &=& \frac{\l_1}{2}|\phi_1|^4 + \frac{\l_2}{2}|\phi_2|^4
 + \l_3 |\phi_1|^2 |\phi_2|^2 
+ \l_4 |\phi_1^\dagger \phi_2|^2 + \frac{\l_5}{2} [(\phi_1^\dagger \phi_2)^2 + h.c.],\\
V_4^b &=& \frac{\s_1}{2}|\eta_1|^4 + \frac{\s_2}{2}|\eta_2|^4
 + \s_3 |\eta_1|^2 |\eta_2|^2   
+ \s_4 |\eta_1^\dagger \eta_2|^2 + \frac{\s_5}{2} [(\eta_1^\dagger \eta_2)^2 + h.c.],\\
V_4^c &=& ‎‎\sum_{i,j=1,2} \Big[ \l_{3ij}|\phi_i|^2 |\eta_j|^2 + \l_{4ij}|\phi_i^\dagger \eta_j|^2
+ \frac{\l_{5ij}}{2} [(\phi_i^\dagger \eta_j)^2 + h.c.] \Big], \\
V &=& V_2 + V_4^a + V_4^b + V_4^c.
\eea
\eesub
Here, $V_2$ and $V_4^{a,b,c}$ respectively contain the quadratic and quartic terms of the scalar potential.
All parameters in $V$ are chosen real to avoid introducing CP-violation. The discrete $\mathbb{Z}_2 \times \mathbb{Z}^\prime_2$ symmetry prevents $\eta_{1,2}$ from picking up vacuum expectation values (VEVs). However, $\phi_{1,2}$ can pick up VEVs 
$v_{1,2}$. The doublets can therefore be expressed as
\begin{eqnarray}
\phi_i = \begin{pmatrix}
\phi_i^+ \\
\frac{1}{\sqrt{2}} (v_i + h_i + i z_i)
\end{pmatrix},~  
\eta_i = \begin{pmatrix}
H_i^+ \\
\frac{1}{\sqrt{2}} (H_i + i A_i)
\end{pmatrix}, ~i = 1,2.
\end{eqnarray} 
This particular assignment of the discrete charges ensures that $\eta_1$ and $\eta_2$ neither mix with each other nor with $\phi_1$ and $\phi_2$. Therefore, the component scalars of $\eta_i$, i.e. $H^+_i,H_i,A_i$, are mass eigenstates themselves with masses $M_i^+, M_{H_i}$ and $M_{A_i}$ respectively. Moreover, the discrete symmetry renders the lightest of each component scalar of $\eta_1$ and $\eta_2$ completely stable. The CP-even components $H_{1,2}$ therefore are possible DM candidates. The doublets $\eta_{1,2}$ are thus \emph{inert} having the following mass spectrum.
\besub
\bea
M^2_{H_{j}} &=& \mu_1^2 + \frac{1}{2}(\l_{31j} + \l_{41j} + \l_{51j})v_1^2
+ \frac{1}{2}(\l_{32j} + \l_{42j} + \l_{52j})v_2^2	\\
M^2_{A_{j}} &=& \mu_1^2 + \frac{1}{2}(\l_{31j} + \l_{41j} - \l_{51j})v_1^2
+ \frac{1}{2}(\l_{32j} + \l_{42j} - \l_{52j})v_2^2	\\
M^2_{H^+_{j}} &=& \mu_1^2 + \frac{1}{2}\l_{31j} v_1^2
+ \frac{1}{2}\l_{32j}v_2^2
\eea
\eesub
On the contrary, the scalars coming from $\phi_1$ and $\phi_2$ as dictated by $V_1$. The scalars in the "gauge basis" are rotated into the "mass basis" by the action of two mixing angles $\a$ and $\b$ where tan$\beta = \frac{v_2}{v_1}$. The mass eigenstates are therefore the CP-even scalars $h,H$, the CP-odd scalar $A$ and the charged scalar $H^+$. This part is exactly same as in an ordinary 2HDM and we discuss no further on it for brevity. We reiterate that $V_4^a$ describes interaction amongst the \emph{non-inert} scalars coming from $\phi_{1,2}$, $V_4^b$ parametrises interaction of the inert scalar sector with the non-inert one, and, $V_4^c$ encodes interactions within the inert sector. 

A counting of the independent model parameters is in order. In the active sector, the relevant parameters are
$\{m_{11},m_{22},m_{12},\l_{1-5},
\text{tan}\beta \}$. Of these, $m_{11}$ and $m_{22}$ can be eliminated by using the tadpole conditions. In addition, the quartic couplings $\l_{1-5}$ can be traded off for the physical masses $M_{h,H,A,H^+}$ and the mixing angle $\a$. The independent parameters in the active sector are therefore $\{m_{12},M_h,M_H,M_A,
M_{H^+},\a,\text{tan}\beta\}$. The inert sector is more elaborate. A counting of parameters yields 19 parameters, i.e., $\{\mu_i,\l_{3ij},\l_{4ij},\l_{5ij},\sigma_{1-5}\}$ for $i,j=1,2$. However, some of these can be replaced by the physical masses of the inert scalars and $\l_{Lij} \equiv \l_{3ij} + \l_{4ij} + \l_{5ij}$. Here, $\l_{Lij}$ parameterise the $H_1-H_1-h$, $H_2-H_2-h$, $H_1-H_1-H$ and $H_2-H_2-H$ portal couplings and hence are of paramount importance in a scalar DM scenario such as the present one. In all, we deem the following parameters as our basis set: $\{M_{H_1}, M_{A_1}, M^+_1, M_{H_2}, M_{A_2}, M^+_2, \l_{L11}, \l_{L12}, \l_{L21}, \l_{L22}, \l_{411}, \l_{412}, \l_{511}, \l_{512}, \s_1, \s_2, \s_3, \s_4, \s_5 \}$. The dimensionful parameters 
$\mu_{1,2}$ and the rest of the quartic couplings can be expressed in terms of the basis parameters as 
\besub
\bea
\mu_1^2 &=& M^2_{H_1} - \frac{1}{2}\l_{L11} v^2 c^2_\b
 - \frac{1}{2}\l_{L21} v^2 s^2_\b, \\
\mu_2^2 &=& M^2_{H_2} - \frac{1}{2}\l_{L12} v^2 c^2_\b
 - \frac{1}{2}\l_{L22} v^2 s^2_\b, \\
\l_{421} &=& \frac{M^2_{H_1} + M^2_{A_1} - 2 (M^+_1)^2 
- \l_{411} v^2 c^2_\b}{v^2 s^2_\b}, \\
\l_{521} &=& \frac{M^2_{H_1} - M^2_{A_1} 
- \l_{511} v^2 c^2_\b}{v^2 s^2_\b}, \\
\l_{422} &=& \frac{M^2_{H_2} + M^2_{A_2} - 2 (M^+_2)^2 
- \l_{412} v^2 c^2_\b}{v^2 s^2_\b}, \\
\l_{522} &=& \frac{M^2_{H_2} - M^2_{A_2} 
- \l_{512} v^2 c^2_\b}{v^2 s^2_\b}, \\  
\l_{311} &=& \l_{L11} - \l_{411} - \l_{511}, \\
\l_{321} &=& \l_{L21} - \l_{421} - \l_{521}, \\
\l_{312} &=& \l_{L12} - \l_{412} - \l_{512}, \\
\l_{322} &=& \l_{L22} - \l_{422} - \l_{522}.
\eea
\eesub

\section{One-loop form factors for $H^+ \rightarrow W^+ Z $}\label{form_factor}
 
\begin{figure}
\centering 
\includegraphics[height = 4.0 cm, width = 9 cm]{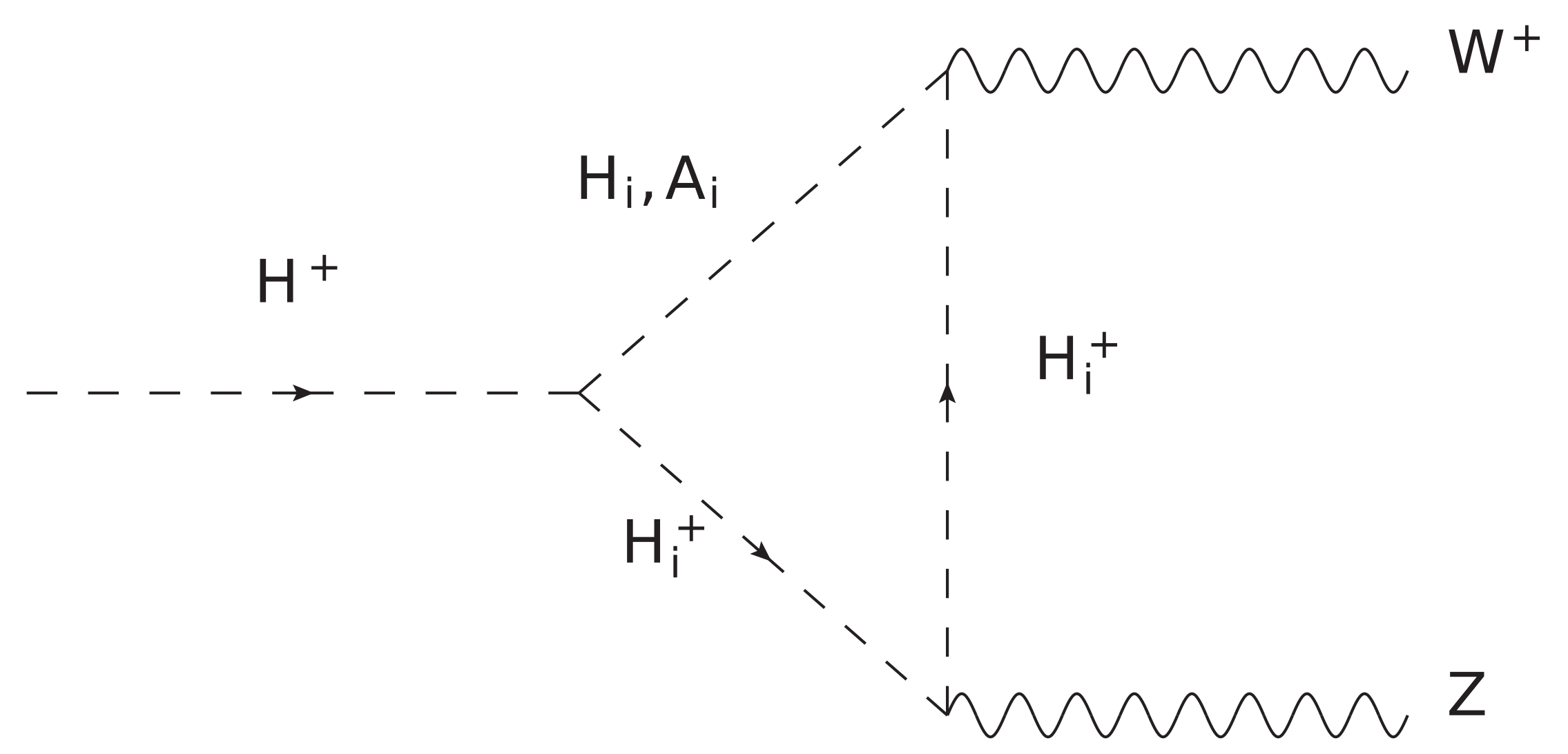}~~
\includegraphics[height = 4.0 cm, width = 9 cm]{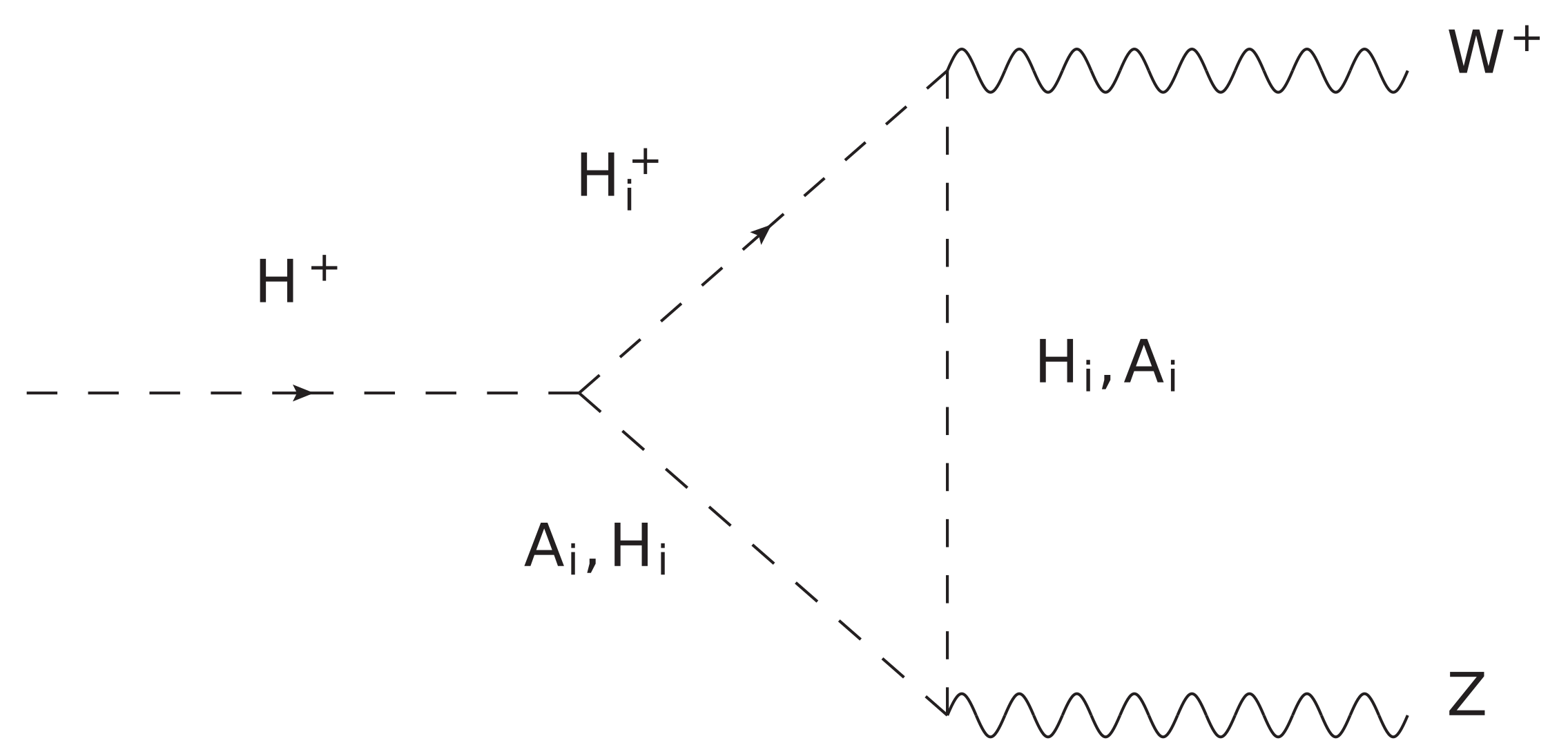}
~~
\includegraphics[height = 4.0 cm, width = 9 cm]{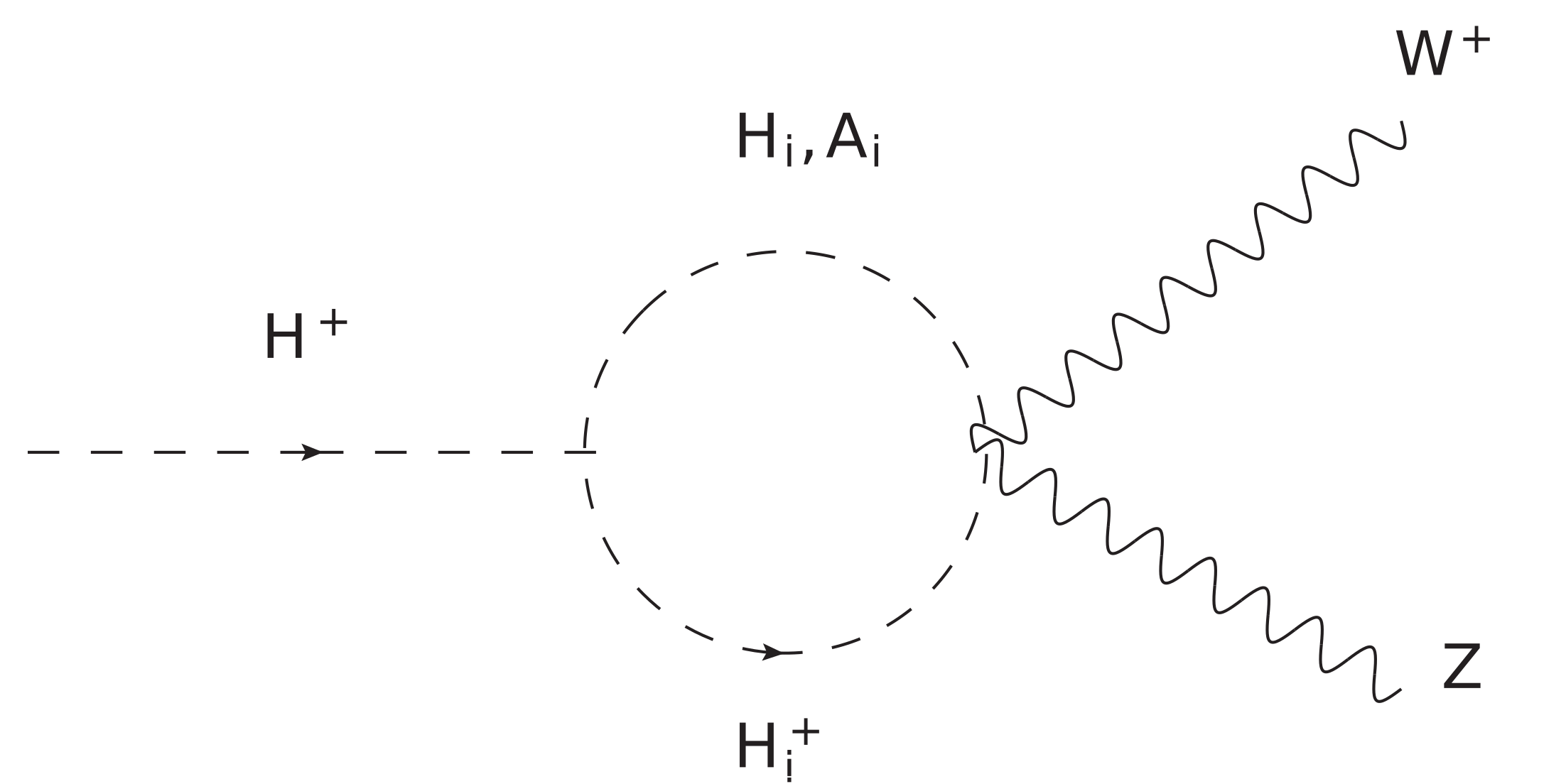}
\caption{One-loop diagrams in set A.}
\label{Fig:setA}
\end{figure}

\begin{figure}
\centering 
\includegraphics[height = 4.0 cm, width = 9 cm]{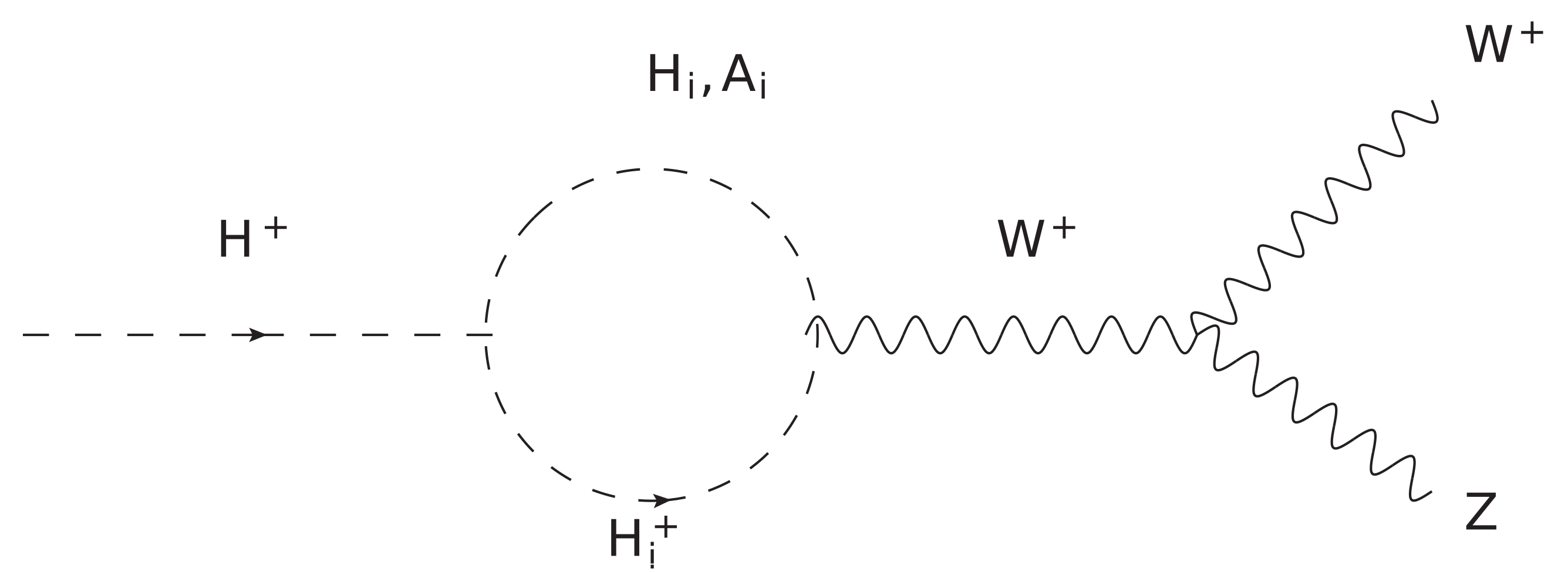}
\caption{One-loop diagrams in set B.}
\label{Fig:setB}
\end{figure}

The amplitude for the $H^+ \rightarrow W^+ Z$ process can be written as
\begin{eqnarray}
i\mathcal{M}(H^+ \rightarrow W^+ Z) = i g m_W V_Z^{\mu \nu} \epsilon_{W \mu}^* (p_W)\epsilon_{Z \nu}^* (p_Z) \,.
\label{eq:1}
\end{eqnarray}
where,
\begin{eqnarray}
V_Z^{\mu \nu} = g^{\mu \nu} F_Z + \frac{p_Z^\mu p_W^\nu}{m_W^2} G_Z + i \epsilon^{\mu \nu \rho \sigma} 
\frac{p_{Z \rho} p_{W \sigma}}{m_W^2} H_Z \,.
\label{eq:2}
\end{eqnarray}
Here $p_W^\mu p_Z^\nu$ denote the incoming momenta of $W^\pm$ and $Z$ and $F_Z,~G_Z$ and 
$H_Z$ are the corresponding form-factors. 
Scalars coming from both the \emph{visible} and \emph{inert} sectors contribute to $H^+ \rightarrow W^+ Z$. A one-loop form factor is therefore a sum of the contributions coming from the two sectors.
\bea
X_Z &=& X_{Z,\text{2HDM}} + X_{Z,\eta_i},
\eea
for $X = F,~G,~H$.
We adopt the $\beta - \alpha = \frac{\pi}{2}$ limit (known as the \emph{alignment limit} in a 2HDM) where the interactions of $h$ to the fermions and gauge bosons
become identical to the corresponding SM values. At this point, the following nonlinear gauge-fixing functions \cite{Fujikawa:1973qs,Bace:1975qi,Gavela:1981ri,Monyonko:1983fi,Monyonko:1986hy,
Hernandez:1999xn,HernandezSanchez:2004tq} are introduced: 
\besub
\bea
f^+ &=& \Big(D^e_\mu + \frac{i g s_W^2}{c_W} Z_\mu \Big) W^{+ \mu}
 - i \xi M_W G^+, \\
f^Z &=& \partial_\mu Z^\mu - \xi M_Z G^0, \\
f^A &=& \partial_\mu A^\mu. 
\eea
\eesub
with $D^e_\mu$ the electromagnetic covariant derivative and $\xi$ the gauge parameter. The corresponding gauge fixing Lagrangian reads:
\besub
\bea
\mathcal{L}_{GF} &=& -\frac{1}{\xi}f^+ f^- - \frac{1}{2\xi}(f^Z)^2 - \frac{1}{2\xi}(f^A)^2
\eea
\eesub
The advantage of using this particular gauge-fixing procedure is that it eliminates the unphysical
$G^+ W^- V$ vertices that arise in the Higgs kinetic-energy sector. 
In addition, for $\xi = 1$, the $Z_\mu (k) W^+_\nu (p) W^-_\rho (q)$ (all momenta incoming) triple-gauge vertex have the following modified Feynman rules.
\besub
\bea
\Gamma^{Z W^+ W^-}_{\rho \nu \mu}(k,p,q) &=& -i g c_W\{g_{\mu\nu}(k - p + \frac{s_W^2}{c_W^2}q)_\rho
 + g_{\nu \rho}(p - q)_\mu  \nonumber \\
 && + g_{\rho \mu}(q - k - \frac{s_W^2}{c_W^2}p)_\nu\} \,.  
\eea
\eesub
With the $G^+ W^- Z$ vertex now absent, the number of diagrams reduces. The diagrams with inert scalars circulating in the loop can then be partitioned into UV-finite sets, say, A and B. The diagrams corresponding to the sets A and B are shown in Fig.\ref{Fig:setA} and Fig.\ref{Fig:setB} respectively. The contribution to 
$F_Z$ coming from $\eta_i$ for set A (set B) is denoted by $F^A_{Z,\eta_i}$ ($F^B_{Z,\eta_i}$). Set B does not contribute to $G_Z$. We denote
\besub
\bea
F_{Z,\text{inert}} &=& \sum_{i=1,2} F^A_{Z,\eta_i} + F^B_{Z,\eta_i}, \\
G_{Z,\text{inert}} &=& \sum_{i=1,2} G^A_{Z,\eta_i}.
\eea
\eesub
 
The decay width of $H^+ \rightarrow W^+ Z$ is given by
\bea
\Gamma (H^+ \rightarrow W^+ Z) = M_{H^+} \frac{\sqrt{\lambda(1,\omega, z)}}{16 \pi} \sum_{i = L, T} |M_{ii}|^2\,,
\eea
where $i = L(T)$ represents the longitudinal and transverse polarization
and,
$\lambda(a,b,c) = (a-b-c)^2 - 4abc , ~\omega = (\frac{M_W}{M_{H^+}})^2, 
z = (\frac{M_Z}{M_{H^+}})^2$. 

The longitudinal and transverse contributions can be expressed in terms of $F_Z,G_Z,H_Z $ as
\besub
\bea
|M_{LL}|^2 &=& \frac{g^2}{4 z}|(1-\omega- z)F_Z + \frac{\lambda(1,\omega, z)}{2 \omega} G_Z|^2 \ \label{Msq1}, \\
|M_{TT}|^2 &=& g^2 (2 \omega |F_Z|^2 + \frac{\lambda(1,\omega, z)}{2 \omega} |H_Z|^2 ) \,.
\label{Msq2}
\eea
\eesub

\section{constraints}\label{constraints}
We discuss the relevant constraints on both the scenarios in this section

\subsection{Theoretical constraints}

The scalar potential is subject to the theoretical requirements of perturbativity, unitarity and vacuum stability. A perturbative theory demands that all the quartic couplings remain $\leq 4 \pi$. Therefore, for the more general two-inert doublet case, maintaining 
$|\l_{1-5}|,|\sigma_{1-5}|,|\l_{3ij}|,|\l_{4ij}|,
|\l_{5ij}| \leq 4 \pi$ ensures perturbativity.

Unitarity dictates that the tree-level 
$2 \to 2$ scattering matrix computed between various two particle 
states consisting of charged and neutral scalars \cite{PhysRevD.16.1519} must have eigenvalues bounded from above at $\leq 8\pi$ in magnitude. These eigenvalues are combinations of the quartic couplings. Therefore, demanding unitarity is tantamount to deriving upper bounds on the quartic couplings.

In addition, the following conditions for the two-inert doublet case ensure that the scalar potential remains bounded from below (BFB) along various directions in field space:
\besub
\bea
\l_1 > 0,~\l_2 > 0,~\l_3 + \sqrt{\l_1 \l_2} > 0, 
~\l_3 + \l_4 - |\l_5| + \sqrt{\l_1 \l_2} > 0, \\
\s_1 > 0,~\s_2 > 0,~\s_3 + \sqrt{\s_1 \s_2} > 0, 
~\s_3 + \s_4 - |\s_5| + \sqrt{\s_1 \s_2} > 0. \\
\l_{311} + \sqrt{\l_1 \s_1} > 0,~\l_{311} + \l_{411} - 
|\l_{511}| + \sqrt{\l_1 \s_1} > 0, \\
\l_{312} + \sqrt{\l_1 \s_2} > 0,~\l_{312} + \l_{412} - 
|\l_{512}| + \sqrt{\l_1 \s_2} > 0, \\
\l_{321} + \sqrt{\l_2 \s_1} > 0,~\l_{321} + \l_{421} - 
|\l_{521}| + \sqrt{\l_2 \s_1} > 0, \\
\l_{322} + \sqrt{\l_2 \s_2} > 0,~\l_{322} + \l_{422} - 
|\l_{522}| + \sqrt{\l_2 \s_2} > 0.
\eea
\eesub  
The corresponding conditions for the single inert doublet scenario are obtained by omitting the inequalities involving $\s_{2-5},\l_{312},\l_{412},\l_{512},\l_{322},\l_{422},\l_{522}$.

\subsection{Oblique parameters}
The additional scalars modify the oblique parameters \cite{Peskin:1991sw} with respect to the SM value. Moreover in multi-doublet scenarios~\cite{Grimus:2008nb}, the $T$-parameter is the most constraining
among the oblique parameters that restricts scalar mass splittings. The total $T$-parameter can be written in terms of the SM and BSM contribution 
($\Delta T$) as:
\bea
T &=& T_{\rm SM} + \Delta T \,.
\eea
The most updated bound on the New Physics contribution to $T$-parameter is \cite{pdg}:
\bea 
\Delta T &=& 0.07 \pm 0.12. 
\eea
In our case, the source of BSM contribution is two-fold, \textit{i.e.}, contribution arising from 2HDM and the scalar octet. Thus,
\besub
\bea
\Delta T &=& T_{\rm 2HDM} + T_{\eta_1} + T_{\eta_2}, \\
T_{\rm 2HDM} &=& \frac{1}{16 \pi s^2_W M^2_W}
\Big[F(M^2_{H^+},M^2_A) + s^2_{\b-\a}\Big(F(M^2_{H^+},M^2_H) - F(M^2_H,M^2_A)\Big) \nonumber \\
&&
+ c^2_{\b-\a}\Big(F(M^2_{H^+},M^2_h) - F(M^2_A,M^2_h)
+ F(M^2_W,M^2_H) - F(M^2_W,M^2_h) \nonumber \\
&&
+ F(M^2_Z,M^2_h) - F(M^2_Z,M^2_H)
+ 4 M^2_Z \bar{B_0}(M^2_Z,M^2_H,M^2_h)
- 4 M^2_W \bar{B_0}(M^2_W,M^2_H,M^2_h)\Big)\Big] \,, \label{T2HDM} \nonumber \\
T_{\eta_i} &=& \frac{1}{16 \pi s^2_W M^2_W}
\Big[F((M^+_i)^2,M^2_{H_i}) + F((M^+_i)^2,M^2_{A_i}) 
 - F(M^2_{H_i},M^2_{A_i})\Big] \label{TS} \,,
\eea
\eesub
where,
\besub
\bea
F(x,y) &=&  \frac{x+y}{2} - \frac{xy}{x-y}~{\rm ln} \bigg(\frac{x}{y}\bigg)~~~ {\rm for} ~~~x \neq y \,, \nonumber \\
&=& 0~~~ {\rm for} ~~~ x = y \, \\
\bar{B_0}(m_1^2,m_2^2,m_3^2) &=& \frac{m_1^2 \text{log} (m_1^2) - m_3^2 \text{log} (m_3^2)}{m_1^2 - m_3^2}
- \frac{m_1^2 \text{log} (m_1^2) - m_2^2 \text{log} (m_2^2)}{m_1^2 - m_2^2}. 
\eea
\eesub

\subsection{$h \to \g \g$ signal strength} For $\b - \a = \frac{\pi}{2}$, the only signal strength to get modified \emph{w.r.t.} the SM value is for the $h \to \g \g$ channel. The physical charged scalars $H^+, H^+_{1,2}$ modify the diphoton amplitude as  
\besub
\bea
\mathcal{M}_{h \to \gamma \gamma} &=&
\mathcal{M}^{\text{SM}}_{h \to \gamma \gamma} 
+ \mathcal{M}^{\prime}_{h \to \gamma \gamma} \\
\mathcal{M}^{\text{SM}}_{h \to \gamma \gamma} &=& 
\sum_f N_f Q_f^2 A_{1/2}\Big(\frac{M^2_\phi}{4 M^2_f}\Big)
 + A_1\Big(\frac{M^2_h}{4 M^2_W}\Big), \\
\mathcal{M}^{\prime}_{h \to \gamma \gamma} &=& 
\frac{\l_{h H^+ H^-} v}{2 M^2_{H^+}} A_0\Big(\frac{M^2_h}{4 M^2_{H^+}}\Big)
+ ‎‎\sum_{i=1,2} \frac{\l_{h H_i^+ H_i^-} v}{2 M^2_{H_i^+}} A_0\Big(\frac{M^2_\phi}{4 M^2_{H_i^+}} \Big) \,, \\ 
\Gamma_{h \to \gamma \gamma} &=& \frac{G_F \a^2 M_h^3}{128 \sqrt{2} \pi^3} |\mathcal{M}_{h \to \gamma \gamma}|^2.
\label{htogaga}
\eea
\eesub
where $G_F$ and $\a$ denote respectively the Fermi constant and the QED fine-structure constant. The loop functions are listed below.
\besub
\bea
A_{1/2}(x) &=& \frac{2}{x^2}\big((x + (x -1)f(x)\big), \\
A_1(x) &=& -\frac{1}{x^2}\big((2 x^2 + 3 x + 3(2 x -1)f(x)\big), \\
A_0(x) &=& -\frac{1}{x^2}\big(x - f(x)\big),  \\
\text{with} ~~f(x) &=& \text{arcsin}^2(\sqrt{x}); ~~~x \leq 1 
\nonumber \\
&&
= -\frac{1}{4}\Bigg[\text{log}\frac{1+\sqrt{1 - x^{-1}}}{1-\sqrt{1 - x^{-1}}} -i\pi\Bigg]^2; ~~~x > 1.
\eea
\eesub

where $A_{1/2}(x), A_1(x)$ and 
$A_0(x)$ are the respective amplitudes for the spin-$\frac{1}{2}$, spin-1 and spin-0 particles in the loop.
The latest 13 TeV results on the diphoton signal strength from ATLAS~\cite{Aaboud:2018xdt} and CMS~\cite{Sirunyan:2018ouh} read
\besub
\bea
\mu_{\g\g} &=& 0.99^{+0.14}_{-0.14}, \\
&=& 1.18^{+0.17}_{-0.14}.
\eea
\eesub
Upon using the standard combination of signal strengths and uncertainties, we obtain $\mu_{\g\g} \simeq 1.06 \pm
0.1$ and impose this constraint at 2$\sigma$.

\subsection{Dark matter constraints}

In concurrence with the observation on the relic abundance of the dark matter provided the PLANCK experiment, we demand for our model
\bea
0.117 \leq \Omega h^2 \leq 0.123.
\eea
Further, the dark matter parameter space is constrained significantly by the direct
detection experiments such as LUX~\cite{Akerib:2016vxi}, PandaX-II~\cite{Zhang:2018xdp} and Xenon-1T~\cite{Aprile:2018dbl}. We abide by the most stringent bound that comes from Xenon-1T, in the present analysis.

\section{Dark matter phenomenology} \label{DM}

The present framework comprises two different dark sectors endowed by two different discrete symmetries. Two different DM candidates $H_1,~H_2$, one from each sector, thus emerge. The total relic density is thus the sum of the individual components, i.e., $\Omega h^2 = \Omega_1 h^2 + \Omega_2 h^2$. Here, $h$ refers to the reduced Hubble parameter. The individual relic densities are determined by solving the Boltzmann equations. Before 
writing the equations themselves, we define the parameter
$x = \frac{\mu}{T}$, where $\mu$ is the reduced mass
defined through $\mu = \frac{M_{H_1} M_{H_2}}
{M_{H_1} + M_{H_2}}$. The comoving number density for
$H_{1,2}$ at a temperature $T$ is denoted by 
$Y_{1,2}(x)$. In terms of these, the Boltzmann equations are given by 
\besub
\bea
\frac{d y_1}{dx} &=& -\frac{1}{x^2} \Bigg[ \sum_{B} \bra \s v \ket_{H_1 H_1 \to B B} \big(y^2_1 - (y_1^{\text{eq}})^2 \big) \Theta(M_{H_1} - M_\phi) \nonumber \\
&&
+ \bra \s v \ket_{H_1 H_1 \to H_2 H_2} 
\Bigg( y_1^2 - \frac{(y_1^{\text{eq}})^2}{(y_2^{\text{eq}})^2} y_2^2 \Bigg) 
\Theta(M_{H_1} - M_{H_2}) \nonumber \\
&&
- \bra \s v \ket_{H_2 H_2 \to H_1 H_1} 
\Bigg( y_2^2 - \frac{(y_2^{\text{eq}})^2}{(y_1^{\text{eq}})^2} y_1^2 \Bigg) 
\Theta(M_{H_2} - M_{H_1}) \Bigg], \\
\frac{d y_2}{dx} &=& -\frac{1}{x^2} \Bigg[ \sum_{B} \bra \s v \ket_{H_2 H_2 \to B B} \big(y^2_2 - (y_2^{\text{eq}})^2 \big) \Theta(M_{H_2} - M_\phi) \nonumber \\
&&
+ \bra \s v \ket_{H_2 H_2 \to H_1 H_1} 
\Bigg( y_2^2 - \frac{(y_2^{\text{eq}})^2}{(y_1^{\text{eq}})^2} y_1^2 \Bigg) 
\Theta(M_{H_1} - M_{H_2}) \nonumber \\
&&
- \bra \s v \ket_{H_1 H_1 \to H_2 H_2} 
\Bigg( y_1^2 - \frac{(y_1^{\text{eq}})^2}{(y_2^{\text{eq}})^2} y_2^2 \Bigg) 
\Theta(M_{H_1} - M_{H_2}) \Bigg].
\eea
\eesub 
In the above, $B$ denotes a bath particle 
and $y_i \equiv 0.26 M_{\text{Pl}} \sqrt{g_*} \mu Y_i$.
The equilibrium density $Y_i^{\text{eq}}$ can now be expressed as
\bea
Y_i^{eq}(x) &=& 0.145 \frac{g}{g_*} x^{3/2} \Bigg(\frac{M_{H_i}}{\mu} \Bigg)^{3/2} 
e^{-x \Big( \frac{M_{H_i}}{\mu} \Big)}
\eea

\begin{figure}
\centering 
\includegraphics[height = 4.0 cm, width = 5 cm]{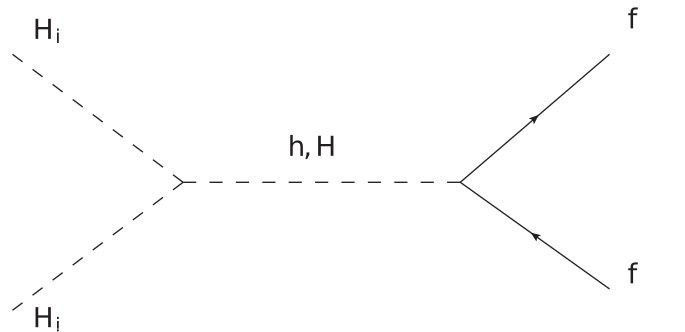}~~~
\includegraphics[height = 4.0 cm, width = 5 cm]{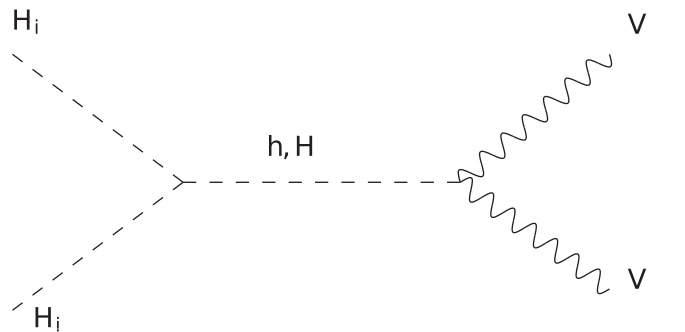}~~~
\includegraphics[height = 4.0 cm, width = 5 cm]{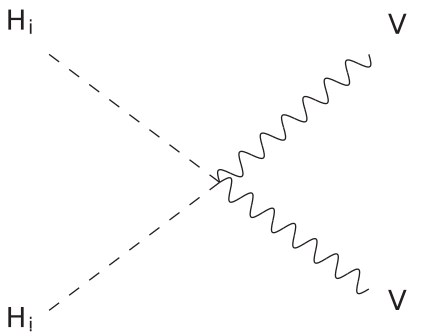}
\\
\includegraphics[height = 4.0 cm, width = 5 cm]{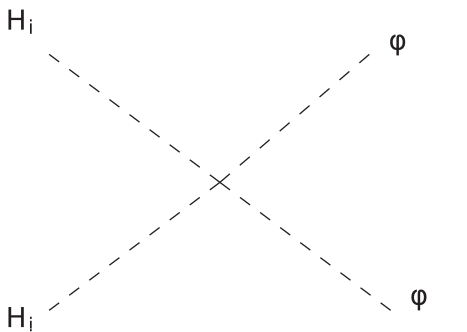}~~~
\includegraphics[height = 4.0 cm, width = 5 cm]{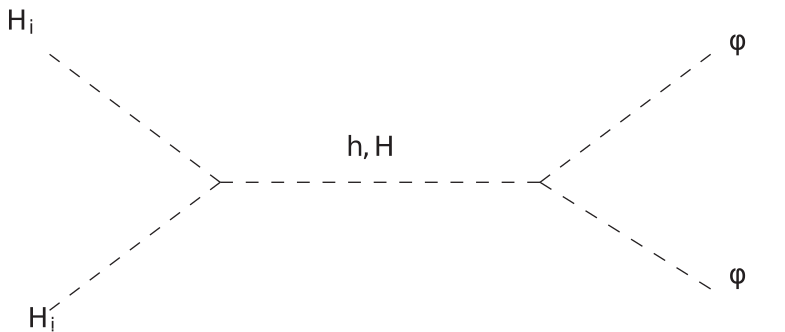}\\
\includegraphics[height = 4.0 cm, width = 5 cm]{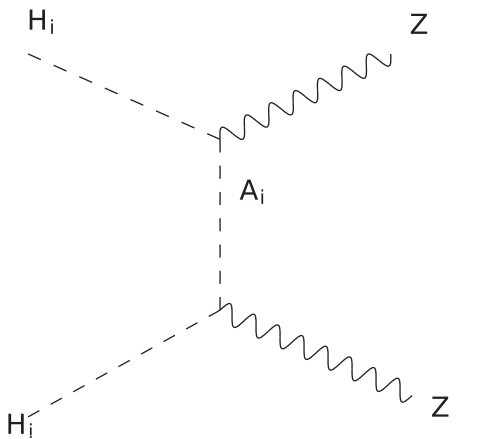}~~~
\includegraphics[height = 4.0 cm, width = 5 cm]{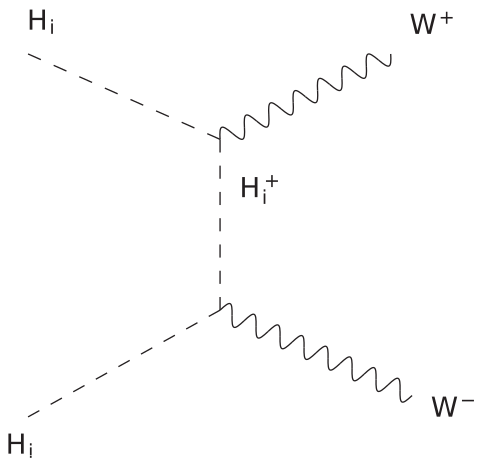}\\
\includegraphics[height = 4.0 cm, width = 5 cm]{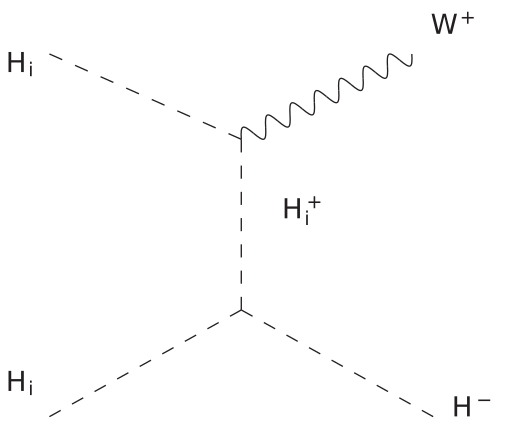}~~~
\includegraphics[height = 4.0 cm, width = 5 cm]{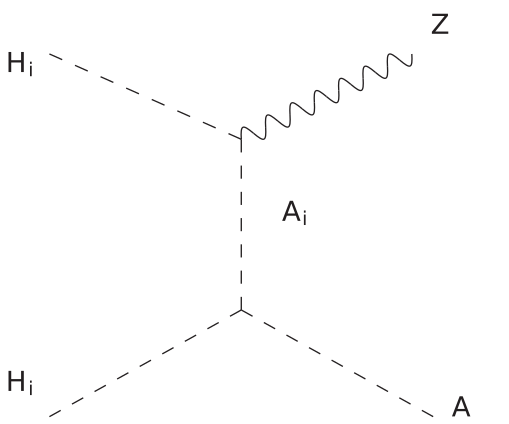}
\caption{DM annihilation diagrams relevant for the [$M_W$,500 GeV] mass range.}
\label{Fig:anni}
\end{figure}

\begin{figure}
\centering 
\includegraphics[height = 4.0 cm, width = 5 cm]{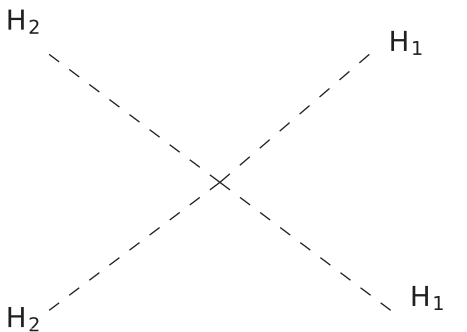}~~~
\includegraphics[height = 4.0 cm, width = 5 cm]{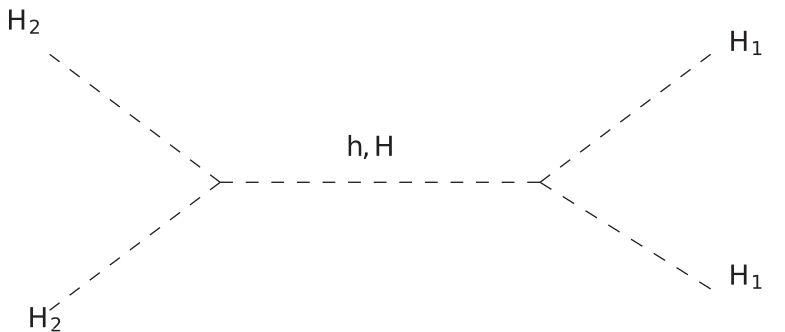}
\caption{DM conversion diagrams for $M_{H_2} > M_{H_1}$.}
\label{Fig:conv}
\end{figure}

Apart from the annihilation processes $H_i H_i \to B B$ (see Fig.\ref{Fig:anni}), assuming $M_{H_2} > M_{H_1}$, 
the $H_2 H_2 \to H_1 H_1$ conversion process (see Fig.\ref{Fig:conv}) forms a key aspect of the analysis as will be elaborated later. Finally, the relic abundance $\Omega_i h^2$ is obtained using
\bea
\Omega_i h^2 &=& \frac{8.54 \times 10^{-13}}{\sqrt{g_*}}
\frac{M_{H_i}}{\mu} y_i\Big( \frac{\mu}{M_{H_i}} x_\infty \Big),
\eea
where $x_\infty$ implies a large enough post-decoupling value.

On another part, SI elastic scatterings involving the DM particles and the nucleons are triggered
by t-channel exchanges of $h$ and $H$. The expression for the SI-DD cross section corresponding to $H_i$ reads:
\bea
\s_i^{\text{SI}} &=& \frac{f_n^2 \mu^2_{i} m^2_n}{4\pi M^2_{H_i}} \Bigg( \frac{\l_{L1i}}{M_h^2} 
+ \frac{\l_{L2i}}{M_H^2} \Bigg)^2, i = 1,2.
\eea 
In the above, $\mu_i = \frac{m_n M_{H_i}}{m_n + M_{H_i}}$
is the DM-nucleon reduced mass and $f_n$ refers to the nuclear form factor. Moreover, In this two-component
DM framework, the effective SI-DD cross sections relevant for each of the candidates can be expressed
by scaling the individual DM-nucleon cross-section by the relative abundance of that particular
component ($\Omega_i h^2$) in total DM relic density 
($\Omega h^2$). That is,
\bea
\sigma^{SI}_{i,~\text{eff}} &=& 
\frac{\Omega_i}{\Omega} \sigma^{SI}_i.
\eea
The effective SI-DD cross sections corresponding to the two DM species must satisfy the XENON 1T bound. 

\begin{figure}
\centering 
\includegraphics[height = 7.0 cm, width = 8 cm]{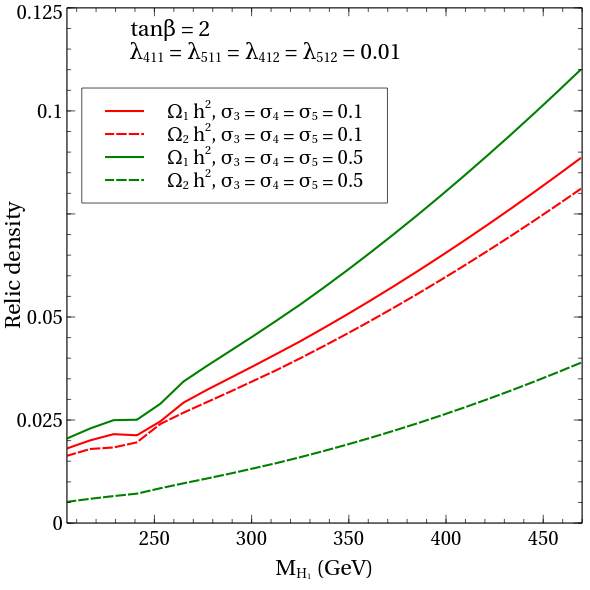}~~~
\includegraphics[height = 7.0 cm, width = 8 cm]{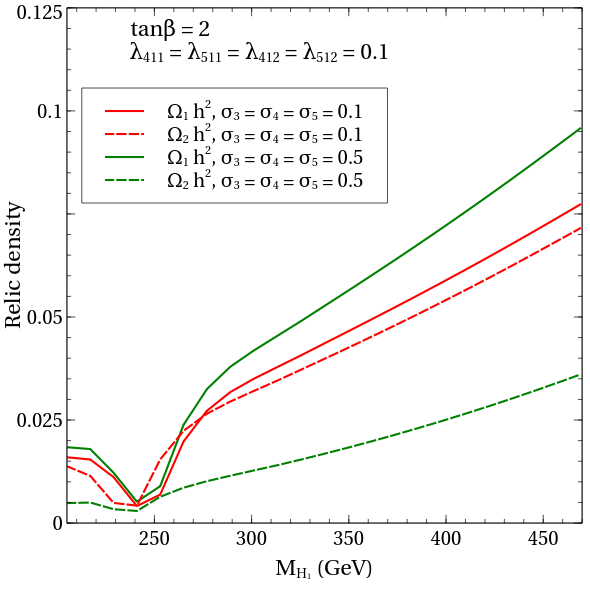}~~~
\caption{Variations of $\Omega_{1,2}h^2$ with DM mass for tan$\beta$ = 2. The left (right) panel corresponds to 
$\l_{411}$ = 0.01 (0.1). Values for the other parameters are given in the text.}
\label{Fig:relic_2_p01}
\end{figure}

In this work, we probe the IDM \emph{desert region} i.e., $M_W < M_{\text{DM}} < 500$ GeV, where a $\simeq$ 0.1 relic is not obtainable in presence of a single inert doublet. Our objective is to find out if the desert region can be revived upon introduction of the second doublet. We first show the variation of the individual relics $\Omega_{1,2}h^2$ versus $M_{H_1}$ in Fig.\ref{Fig:relic_2_p01}. We fix $M_{H_2}-M_{H_1}$ = 10 GeV in order to kinematically allow for the $H_2 H_2 \to H_1 H_1$ conversion process. The conversion amplitude is sensitive to the combination $\s_3 + \s_4 + \s_5$. We set 
$\s_3=\s_4=\s_5$ for simplicity and subsequently take 
$\s_3$ = 0.1, 0.5. The rest of the parameters in the IDM sector are chosen to be: $M_{A_i} - M^+_i = 
M^+_i - M_{H_i}$ = 1 GeV; $\l_{Lij} = 0.01$ for $i,j$ = 1,2; $\l_{411} = \l_{511} = \l_{412} = \l_{512}$ = 0.01, 0.1. The narrow mass splittings in a particular inert are motivated from the IDM itself wherein co-annihilations are triggered for small mass gaps. The small values for 
$\l_{Lij}$ help mitigate the DD bound. The 2HDM spectrum taken is reads $M_H = M_{H^+} = 500$ GeV, $M_A$ = 510 GeV. The left (right) panel of Fig.\ref{Fig:relic_2_p01} corresponds to $\s_3$ = 0.01 (0.1). Upon inspecting the left panel, it is seen that the total relic density attains the requisite value for $M_{H_1} \simeq 385$ GeV in case of $\s_3$ = 0.1. More specifically, $\Omega_1 h^2 \simeq 0.62$ and $\Omega_2 h^2 \simeq 0.55$ at this mass point. It is therefore demonstrated that the current framework can revive the IDM desert region. Even more striking is the observation for $\s_3$ = 0.5. In this case, $H_1$ alone having mass $\simeq$ 465 GeV can account for the observed relic. This enhanced relic of an individual DM component is a direct consequence of an enhanced DM-DM conversion rate. That is, the 
$H_2 H_2 \to H_1 H_1$ rate is higher for $\s_3$ = 0.5 than it is for $\s_3$ = 0.1 and the abundance of $H_1$ is higher in case of the former. The abundance of $H_2$ becomes accordingly lower. 

On another part, an important difference between the current framework and the one in REF is that the annihilation final states in the current scenario can involve 2HDM scalars. An example worthy of elucidation here is $H_i H_i \to W^\pm H^\mp, Z A$ for $i$ = 1, 2.
Since, The edges seen at $M_{H_1} \simeq$ 240 GeV can be understood as the kinematical thresholds for these two final states. Now, the $H_1 H_1 \to W^\pm H^\pm$ channel involves both s- and t-channel amplitudes. The s-channel amplitude involves the $\l_{h H_1 H_1}$ trilinear coupling which can further be expressed in terms of 
$\l_{Lij}$ and 
tan$\beta$ for $i,j$ = 1, 2. On the other hand, the t-channel amplitude involves the $\l_{H^+ H_1^- H_1}$ coupling that in turn depends on the values of $\l_{411}$ and $\l_{511}$. Increasing the value to $\l_{411}$ = $\l_{411}$ = 0.1 enhances the $H_1 H_1 \to W^\pm H^\mp$
annihilation rate thereby leading to a reduced relic. This is concurred by an inspection of the right panel.
The edges at the kinematical threshold is expectedly sharper in this case.
\begin{figure}
\centering 
\includegraphics[height = 7.0 cm, width = 8 cm]{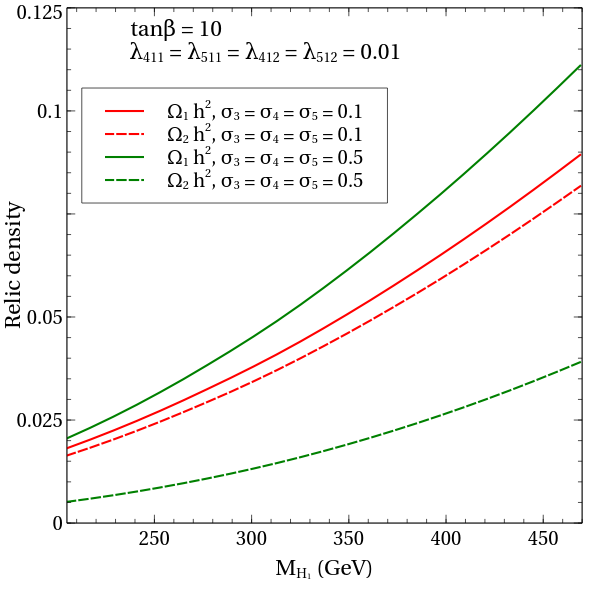}~~~
\includegraphics[height = 7.0 cm, width = 8 cm]{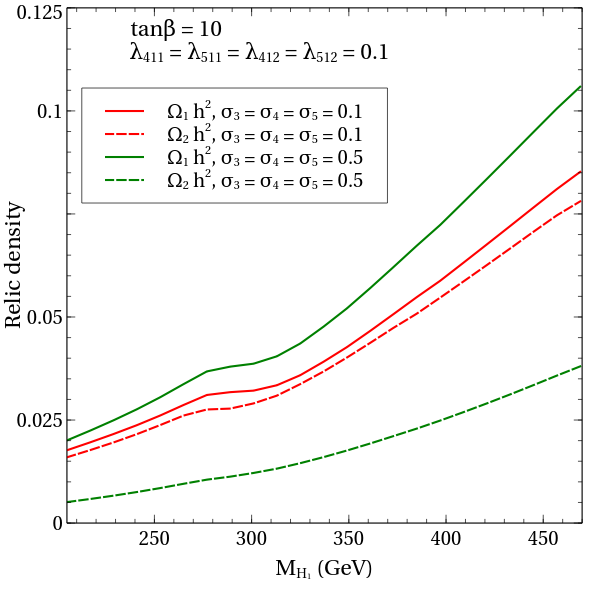}~~~
\caption{Variations of $\Omega_{1,2}h^2$ with DM mass for tan$\beta$ = 10. The left (right) panel corresponds to 
$\l_{411}$ = 0.01 (0.1). Values for the other parameters are given in the text.}
\label{Fig:relic_10_p01}
\end{figure}

Fig.\ref{Fig:relic_10_p01} displays the corresponding curves for tan$\beta$ = 10. The behaviour is qualitatively the same as the case of tan$\beta$ = 2.
However, the magnitude of $\l_{H^+ H_i^- H_i}$ reduces 
accordingly. The edges corresponding to the kinematical 
thresholds are consequently softened.

\begin{figure}
\centering 
\includegraphics[height = 7.0 cm, width = 8 cm]{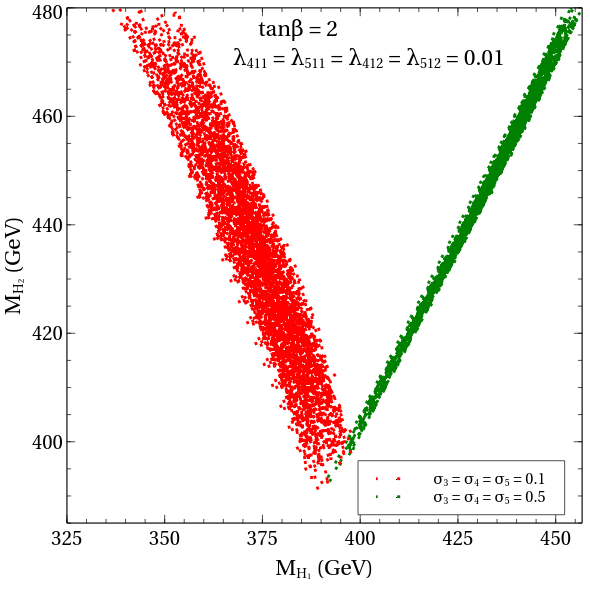}~~~
\includegraphics[height = 7.0 cm, width = 8 cm]{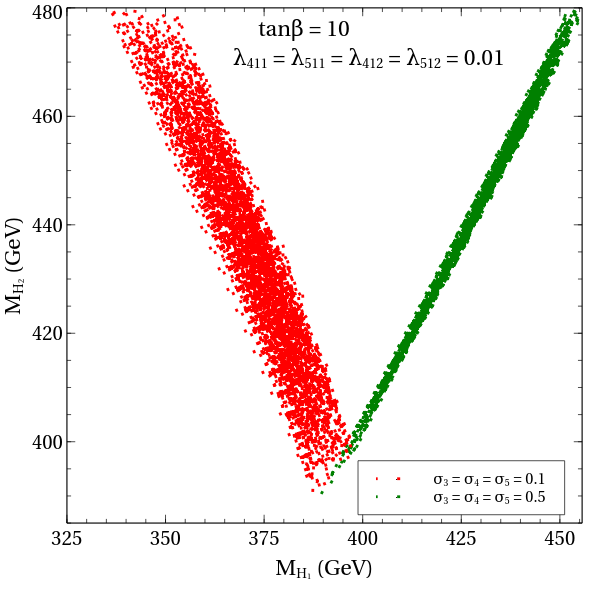}
\caption{Allowed parameter regions in $M_{H_1}-M_{H_2}$ plane.}
\label{Fig:mh1_mh2}
\end{figure}

We show the parameter region allowed the relic and DD constraints in the $M_{H_1}-M_{H_2}$ plane for tan$\beta$ = 2, 10 in Fig.\ref{Fig:mh1_mh2}. The role of the \emph{conversion couplings} $\s_3,\s_4$ and $\s_5$ is clarified here. The kinematics whenever $M_{H_1} = M_{H_2}$ implies little to no conversion regardless of the value of the conversion couplings. This is precisely why the two parameter bands for tan$\beta$ = 2 and 10 meet at the $M_{H_1} \simeq M_{H_2}$ corner. Increasing $M_{H_2} - M_{H_1}$ and keeping the conversion couplings fixed increases the conversion rate. In other words, upon increasing the conversion couplings, demanding the same conversion rate is tantamount to reducing the mass gap between $H_1$ and $H_2$. This explains why the parameter band in the $M_{H_1}-M_{H_2}$ plane shifts towards right
upon increasing $\s_3,\s_4$ and $\s_5$ from 0.1 to 0.5.

\section{Combined results} \label{combined}

We aim to correlate DM phenomenology with the strength of the $H^+ W^- Z$ vertex in this section. Similarly as in the previous section, we take tan$\beta$ = 2, 10; 
$\s_3 = \s_4 = \s_5$ = 0.1, 0.5; $M_i^+ - M_i$ = 1 GeV and $\l_{Lij}$ = 0.01 for 
$i,j$ = 1, 2. We further choose $M_{H_1}$ = 380 GeV and $M_{H_2}$ = 420 GeV, values typical of the desert region.
In addition, the following variation is made.
\besub
\bea
|\l_{411}| = |\l_{412}| \leq 4\pi;~|\l_{511}| = 
|\l_{512}| \leq 4\pi; \\
~0 \leq \Delta M \equiv 
M_{A_1} - M^+_1 = M_{A_2} - M^+_2 \leq 200~\text{GeV}.
\eea
\eesub

The 2HDM contribution is computed taking $M_{H} = M_{H^+}$ = 200 GeV and $M_A$ = 500 GeV. Such a large $M_A - M_{H^+}$ is motivated by the fact that the 2HDM contribution increases with increasing this mass-splitting~\cite{Kanemura:1997ej,Moretti:2015tva,Chakrabarty:2020msl}. We skip the details of the 2HDM contribution for brevity and focus on the one from the inert doublets.

\begin{figure}
\centering 
\includegraphics[height = 7.0 cm, width = 8 cm]{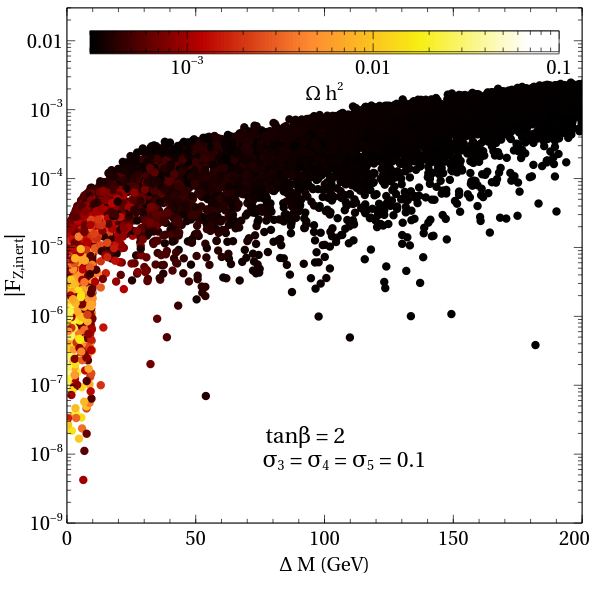}~~~
\includegraphics[height = 7.0 cm, width = 8 cm]{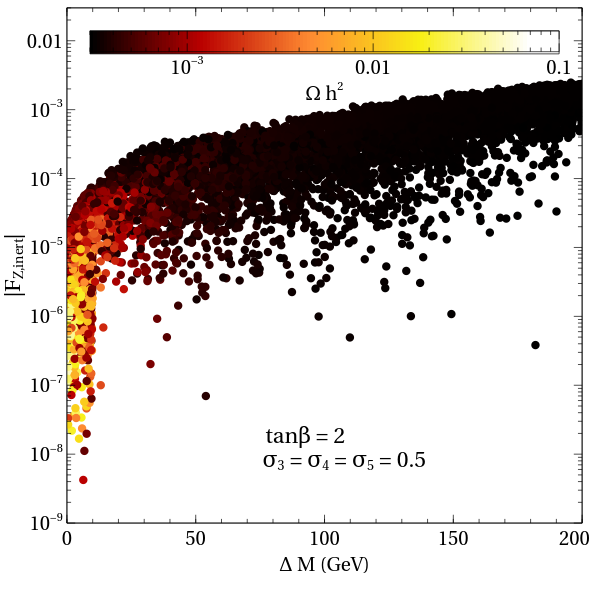}
\caption{Heat plot showing the variation of relic density and $F_{Z,\text{inert}}$ with $\Delta M$ for tan$\beta$ = 2.}
\label{Fig:fz_2}
\end{figure}

\begin{figure}
\centering 
\includegraphics[height = 7.0 cm, width = 8 cm]{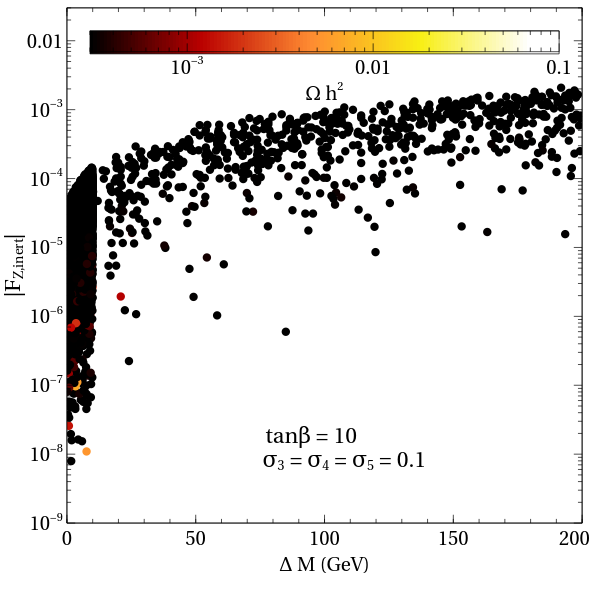}~~~
\includegraphics[height = 7.0 cm, width = 8 cm]{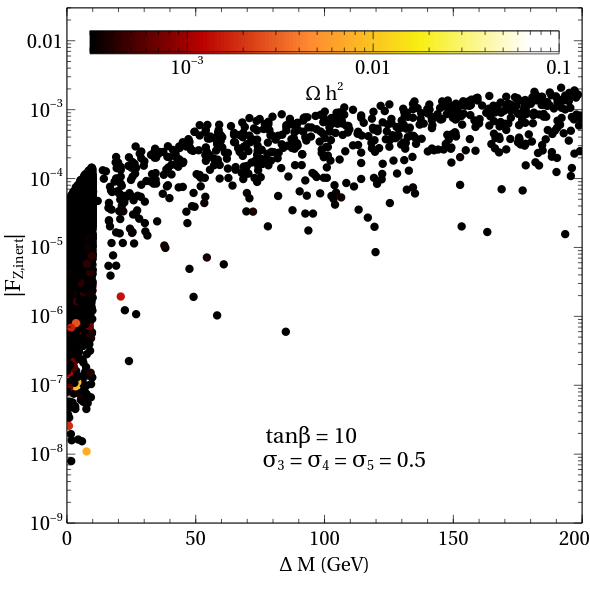}
\caption{Heat plot showing the variation of relic density and $F_{Z,\text{inert}}$ with $\Delta M$ for tan$\beta$ = 10.}
\label{Fig:fz_10}
\end{figure}

Fig.\ref{Fig:fz_2} is a scatter plot in the $\Delta M-F_{Z,\text{inert}}$ plane for tan$\beta$ = 2. It is seen that the form factor increases with increasing $\Delta M$. This can be explained as follows: Increasing $\Delta M$ leads to an increase in the magnitudes of $\l_{421}, \l_{521}, \l_{422}, \l_{522}$. This enhances 
$\l_{H^+ H_i^- H_i}$ and $\l_{H^+ H_i^- A_i}$, and, ultimately $F_{Z,\text{inert}}$. At the highest mass splitting taken, i.e., $\Delta M$ = 200 GeV, $F_{Z,\text{inert}} \simeq 2.4 \times 10^{-3}$ while $F_{Z,2HDM} \simeq 1.7 \times 10^{-2}$ thereby indicating 
$\simeq 15\%$ enhancement due to the inert sector.
The contribution of the inert scalars to $G_Z$ is relatively more suppressed with the inert scalars accounting only for $\simeq 1\%$ of the total contribution even for large mass splittings. Fig.\ref{Fig:fz_10}
corresponds to tan$\beta$ = 10. The form factor $F_{Z,2HDM}$ diminishes since it is proportional to tan$\beta$.
In fact, it is $\simeq 3.4 \times 10^{-3}$ for $\Delta M$ = 200 GeV. The contribution from inert scalars does not shrink much though, it is $\simeq 1.9 \times 10^{-3}$. The relative enhancement coming from the inert sector is therefore $\simeq 3.6\%$. 

On the other hand, the larger is $\Delta M$, the smaller is the relic density. This can be attributed primarily to enhanced annihilations to the $H^\pm W^\mp, A Z$ final states due to the enhanced $\l_{H^+ H_i^- H_i}$ trilinear coupling. A secondary reason is the reduction in the co-annihilation rates due to the large mass gap. This anti-correlation between $F_{Z,\text{inert}}$ and $\Omega h^2$ is clear from the heat plots. Expectedly, there is no appreciable change in relic with change in tan$\beta$.
The conversion couplings $\s_3,\s_4$ and $\s_5$ leave an impact on the heat plots.

\section{Conclusions}\label{conclusion}

In this work, we extended the 2HDM by two additional inert scalar doublets that are charged non-trivially under a $\mathbb{Z}_2 \times \mathbb{Z}_2^\prime$ discrete symmetry. A key part of the ensuing phenomenology is the two-component DM scenario originating from the two inert doublets. We explored the DM phenomenology in detail with emphasis on the $M_W \leq M_{\text{DM}} < 500$ GeV mass range, the well-known  desert region where the IDM fails to score the requisite relic. We showed in this study that the desert region can be revived with the introduction of the second doublet, and, highlighted the role of DM-DM conversions in the process. Moreover, the sensitivity of the thermal relic to the 2HDM spectrum and the parameter tan$\beta$ was also illustrated.

On another part, we computed the strength of the radiatively generated $H^+ W^- Z$ interaction was in the non-linear gauge for the present framework. Particularly, the contribution coming from the inert doublets was detailed. The correlation of the $H^+ W^- Z$ interaction
strength with the relic density in the desert region was studied. It was shown that the interaction strength decreases with increasing relic.

\acknowledgements
NC acknowledges financial support from IISc (Indian Institute of Science) through the C.V.Raman postdoctoral fellowship. NC also sincerely thanks Indrani Chakraborty and Rishav Roshan for their help with some of the figures.

\appendix
\section{Appendix}

This section contains analytical expressions of relevant scalar couplings and form factors.

\subsection{Scalar trilinear couplings}
\besub
\bea
\l_{h H^+ H^-} &=& v\{(-\l_3 c^3_\b + (-\l_1 + \l_4 + \l_5)s^2_\b c_\b) s_\a \nonumber \\
&& + (\l_3 s^3_\b + (\l_2 - \l_4 - \l_5)c^2_\b s_\b) c_\a \} \,, \\ 
\lambda_{H H^+ H^-} &=& \cos\alpha  \{v \sin ^2\beta  \cos~\beta  (\lambda_1 - \lambda_4 - \lambda_5) + \lambda_3 v \cos^3\beta \}  \nonumber \\
 && + \sin \alpha ~ \sin\beta \{v \cos^2 \beta 
   (\lambda_2 - \lambda_4 - \lambda_5) + \lambda_3 v \sin^2\beta\} \\
\l_{h H^+ G^-} &=& \frac{v}{4}\{(\l_2 - \l_3 + \l_4 + \l_5) c_\b c_\a 
+ (-\l_2 + \l_3 + \l_4 + \l_5)c_{3\b}c_\a \nonumber \\
&&
+ (\l_1 - \l_3 + \l_4 + \l_5) s_\b s_\a - (-\l_1 + \l_3 + \l_4 + \l_5)s_{3\b}s_\a\}\\
\l_{H H^+ G^-} &=& \frac{v}{4}\{(\l_2 - \l_3 + \l_4 + \l_5) c_\b s_\a 
+ (-\l_2 + \l_3 + \l_4 + \l_5)c_{3\b}s_\a \nonumber \\
&&
- (\l_1 - \l_3 + \l_4 + \l_5) s_\b c_\a + (-\l_1 + \l_3 + \l_4 + \l_5)s_{3\b}c_\a\} \\
\l_{h H_1 H_1} &=& v\big(\l_{L11}c_\b s_\a - \l_{L21}s_\b c_\a \big), \\
\l_{h H_2 H_2} &=& v\big(\l_{L12}c_\b s_\a - \l_{L22}s_\b c_\a \big), \\
\l_{H H_1 H_1} &=& -v\big(\l_{L11}c_\b c_\a + \l_{L21}s_\b s_\a \big), \\
\l_{H H_2 H_2} &=& -v\big(\l_{L12}c_\b c_\a + \l_{L22}s_\b s_\a \big), \\
\l_{h H^+_1 H^-_1} &=& v\big(\l_{311}c_\b s_\a - \l_{321}s_\b c_\a \big) \\
\l_{h H^+_2 H^-_2} &=& v\big(\l_{312}c_\b s_\a - \l_{322}s_\b c_\a \big) \\
\l_{H^+ H^-_1 H_1} &=& \frac{v}{4}\Big(\l_{411} + \l_{511} - \l_{421} - \l_{521}\Big)\sin 2\b \\
\l_{H^+ H^-_1 A_1} &=& \frac{v}{4}\Big(\l_{411} - \l_{511} - \l_{421} + \l_{521}\Big)\sin 2\b \\
\l_{H^+ H^-_2 H_2} &=& \frac{v}{4}\Big(\l_{412} + \l_{512} - \l_{422} - \l_{522}\Big)\sin 2\b \\
\l_{H^+ H^-_2 A_2} &=& \frac{v}{4}\Big(\l_{412} - \l_{512} - \l_{422} + \l_{522}\Big)\sin 2\b
\eea
\eesub

\subsection{Form factors}
The loop functions have been expressed in terms of Passarino-Veltman functions throughout the analysis~\cite{Passarino:1978jh}. The form-factors for the inert doublet $\eta_i$ ($i$ = 1, 2) read
\besub
\bea
F_{Z,\eta_i}^A &=& \frac{1}{16 \pi^2 v c_W} \Big[\lambda_{H^+ H_i^- H_i}
 [(2 - 4 s_W^2) C_{24}(H_i, H_i^+, H_i^+) - 2 C_{24}(H_i^+, A_i, H_i)
  + s_W^2 B_0(q^2 ; H_i^+, H_i)] \nonumber \\
&&
 + \lambda_{H^+ H_i^+ A_i} [(2 - 4 s_W^2) C_{24}(A_i, H_i^+, H_i^+) - 2 C_{24}(H_i^+, H_i, A_i) + s_W^2 B_0(q^2 ; H_i^+, A_i)]\Big] \,, \\
F_{Z,\eta_i}^B &=& \frac{ s^2_W}{16 \pi^2 v c_W}  
\Big[\l_{H^+ H_i^- H_i} \big(B_0(q^2,H_i^+,H_i) + 2 B_1(q^2,H_i^+,H_i)\big) 
\nonumber \\
&&
+ \l_{H^+ H_i^- A_i} \big(B_0(q^2,H_i^+,A_i) + 2 B_1(q^2,H_i^+,A_i)\big)\Big] \,, \\
G_{Z,\eta_i}^A &=& \frac{ M_W^2}{16 \pi^2 v c_W}\Big[\lambda_{H^+ H_i^- H_i} [(2 - 4 s_W^2) (C_{12} + C_{23})(H_i, H_i^+, H_i^+) - 2 (C_{12} + C_{23})(H_i^+, A_i, H_i) ] \nonumber \\
&& + \lambda_{H^+ H_i^- A_i} [(2 - 4 s_W^2) (C_{12} + C_{23})(A_i, H_i^+, H_i^+) - 2 (C_{12} + C_{23})(H_i^+, H_i, A_i)]\Big] \,,
\eea
\eesub

Similarly, the form factors for the 2HDM can be expressed as 
\besub
\bea
F^{A}_{Z,{\rm 2HDM}} &=& \frac{1}{16 \pi^2 v}[- \lambda_{H H^+ H^-} 
\{\frac{s_W^2}{c_W}~ B_0(q^2,H,H^+) - \frac{2}{c_W}~ C_{24}(H^+,A, H) + \frac{2~ c_{2W}}{c_W} ~C_{24}(H,H^+,H^+)\} \nonumber \\
&& + \lambda_{h H^+ G^-} \{\frac{s_W^2}{c_W}~ B_0(q^2,h,G^+) - \frac{2}{c_W}~ C_{24}(G^+,G_0, h) + \frac{2~ c_{2W}}{c_W} ~C_{24}(h,G^+,G^+)\}]  \\
\\
F^{B}_{Z,{\rm 2HDM}} &=& \frac{s^2_W}{16 \pi^2 v c_W}~ [\lambda_{H H^+ H^-}
\{B_0(q^2,H^+,H) + 2 B_1(q^2,H^+,H)\} \nonumber \\
&& - \lambda_{h H^+ G^-} \{B_0(q^2,G^+,h) + 2 B_1(q^2,G^+,h)\}] 
\,, \nonumber \\
G^{A}_{Z,{\rm 2HDM}} &=& \frac{M_W^2}{16 \pi^2 v} [\lambda_{H H^+ H^-}\{-\frac{2}{c_W}~(C_{12}+C_{23})(H^+,A,H) + \frac{2 c_{2W}}{c_W}~(C_{12}+C_{23})(H,H^+,H^+) \} \nonumber \\ 
&& + \lambda_{h H^+ G^-}\{-\frac{2}{c_W}~(C_{12}+C_{23})(G^+,G^0,h) + \frac{2 c_{2W}}{c_W}~(C_{12}+C_{23})(h,G^+,G^+) \}] \,,
\eea
\eesub

The total form factor is the sum of the individual components.
\besub
\bea
F^{A,F}_{Z,{\rm 2HDM}} &=& \frac{2 N_t}{16 \pi^2 v^2 c_W} [ \nonumber \\
&& M_t^2 \zeta_t~(v_b + a_b)~\{4 C_{24}(t,b,b)-B_0(q^2,t,b) - B_0(p_W^2,b,t) - (2 M_b^2 - M_Z^2)~C_0(t,b,b)\} \nonumber \\
&& - M_b^2 \zeta_b~(v_b + a_b)~\{4 C_{24}(t,b,b)-B_0(p_Z^2,b,b) - B_0(q^2,t,b) - (M_t^2 + M_b^2 - M_W^2)~C_0(t,b,b)\} \nonumber \\
&& - M_b^2 \zeta_b~(v_b - a_b)~\{ B_0(p_Z^2,b,b)+ B_0(p_W^2,t,b) + (M_t^2 + M_b^2 - q^2)~C_0(t,b,b)\} \nonumber \\
&& + 2 M_t^2 M_b^2 \zeta_t~(v_b-a_b)~C_0(t,b,b)] + (M_t,\zeta_t,v_b,a_b) \leftrightarrow (M_b,-\zeta_b,v_t,a_t) \,,   \\
F^{B,F}_{Z,{\rm 2HDM}} &=& \frac{4 s_W^2 N_t}{16 \pi^2 v^2 c_W}[M_t^2 \zeta_t~(B_0 + B_1) - M_b^2 \zeta_b~ B_1](q^2, t, b) \,,   \\
G^{A,F}_{Z,{\rm 2HDM}} &=& \frac{4 N_c M_W^2}{16 \pi^2 v^2 c_W} [ M_t^2 \zeta_t~(v_b + a_b) ~(2C_{23}+2C_{12}+C_{11}+C_0) \nonumber \\
&& - M_b^2 \zeta_b~(v_b+a_b)~(2C_{23}+C_{12}) -  M_b^2 \zeta_b~(v_b- a_b)~(C_{12}-C_{11}) ] (t,b,b) \nonumber \\
&& + (M_t,\zeta_t,v_b,a_b) \leftrightarrow (M_b,-\zeta_b,v_t,a_t) \,,
\eea
\eesub

\bea
H^{\rm 1PI}_{Z,F} &=& \frac{4 N_c M_W^2}{16 \pi^2 v^2 c_W} \times \nonumber \\
&& [M_t^2 \zeta_t (v_b + a_b)(C_0 + C_{11}) - M_b^2 \zeta_b (v_b + a_b) C_{12} + M_b^2 \zeta_b (v_b - a_b)(C_{12} - C_{11})] (t,b,b) \nonumber \\
&& + (M_t, \zeta_t, v_b , a_b) \leftrightarrow (M_b, +\zeta_b, v_t , a_t) \,.
\eea

where, 
\bea
v_f = I_f - s_W^2 Q_f , ~ a_f = I_f .
\eea

\besub
\bea
F_Z &=& F_{Z,\eta_i}^{A} + F_{Z,\eta_i}^{B} + F_{Z,{\rm 2HDM}}^{A} + F_{Z,{\rm 2HDM}}^{B}  + F_{Z,{\rm 2HDM}}^{A,F} \,, \\
G_Z &=& G_{Z,\eta_i}^{A} + G_{Z,{\rm 2HDM}}^{A} + G_{Z,{\rm 2HDM}}^{A,F} \,, \\
H_Z &=& H^{\rm 1PI}_{Z,F} \,, \\
F_\gamma &=& F_{\gamma,\eta_i}^{A} + F_{\gamma,\eta_i}^{B} + F_{\gamma,{\rm 2HDM}}^{A} + F_{\gamma,{\rm 2HDM}}^{B} + F_{\gamma,{\rm 2HDM}}^{A,F} \,, \\
G_\gamma &=& G_{\gamma,\eta_i}^{A} + G_{\gamma,{\rm 2HDM}}^{A} + G_{\gamma,{\rm 2HDM}}^{A,F} \,, \\
H_\gamma &=& H^{\rm 1PI}_{\gamma,F} \,.
\eea
\eesub

\bibliography{ref_2IDM} 
\end{document}